\documentclass[12pt]{article}

\usepackage{amsmath}
\usepackage{amssymb}
\usepackage{graphicx}
\usepackage{subfigure}
\usepackage{fancyhdr}
\usepackage{caption}

\newcommand{\citeStrogatz}{[2]} 
\newcommand{\citeLyap}{[3]} 
\newcommand{\citeMCMCMH}{[15]} 
\newcommand{\citeHMC}{[17]} 
\newcommand{\citeHMCO}{[16]} 
\newcommand{\citeHMCall}{[16,17]} 
\newcommand{\citeSlice}{[18]} 
\newcommand{\citesmcphysd}{[25]} 
\newcommand{\citeMCMCbasic}{[10]} 
\newcommand{\citeLyapStrogatz}{[2,3]}
\newcommand{\citeMH}{[28]} 
\newcommand{\citeStatistics}{[8,9]} 

\newcommand{\FigLyap}{\mbox{S1}}
\newcommand{\Figthreepeaks}{\mbox{S2}}
\newcommand{\FigMCMC}{\mbox{S3}}
\newcommand{\Figpineswitz}{\mbox{S4}}
\newcommand{\Figbigsur}{\mbox{S5}}
\newcommand{\Figstarrynight}{\mbox{S6}}

\title{A chaotic dynamical system that paints}

\author
{Tuhin Sahai,$^{1\ast}$ George Mathew,$^{2}$ Amit Surana$^{1}$\\
\\
\normalsize{$^{1}$United Technologies Research Center, East Hartford, CT}\\
\normalsize{$^{2}$United Technologies Research Center, Berkeley, CA}\\
\\
\normalsize{$^\ast$To whom correspondence should be addressed; E-mail:  tuhin.sahai@gmail.com.}
}

\date{}

\begin{document}


\baselineskip18pt


\maketitle


\begin{abstract}
Can a dynamical system paint masterpieces such as Da Vinci's Mona Lisa or Monet's Water Lilies? Moreover, can this dynamical system be chaotic in the sense that although the trajectories are sensitive to initial conditions, the same painting is created every time? Setting aside the creative aspect of painting a picture, in this work, we develop a novel algorithm to reproduce paintings and photographs. Combining ideas from ergodic theory and control theory, we construct a chaotic dynamical system with predetermined statistical properties. If one makes the spatial distribution of colors in the picture the target distribution, akin to a human, the algorithm first captures large scale features and then goes on to refine small scale features.  Beyond reproducing paintings, this approach is expected to have a wide variety of applications such as uncertainty quantification, sampling for efficient inference in scalable machine learning for big data, and developing effective strategies for search and rescue. In particular, our preliminary studies demonstrate that this algorithm provides significant acceleration and higher accuracy than competing methods for Markov Chain Monte Carlo (MCMC).
\end{abstract}


Is it possible to design a dynamical system that paints like a human? Given the availability of efficient modern printing technologies, this may seem like a trivial problem. However, the manner in which a modern printer prints is fundamentally different when compared to a human painting a picture. Roughly speaking, a printer scans each pixel in a pre-determined order and, using a color palette, deposits the appropriate amount of ink for each pixel. In comparison, given the same color palette, a human paints by first capturing the high level (or large scale) features of the picture and then goes on to fill in the low level (or detailed) features. In this paper, we are not attempting to model or mimic the intelligence and creativity of humans when perceiving and painting pictures. Rather, our objective is to design an algorithm that reproduces the human actions of painting by first capturing the large scale features followed by small scale details. Our algorithm is based on the construction of a deterministic (no randomness or noise) dynamical system (described by a set of governing differential equations), which visits states with frequencies prescribed by a user defined distribution.

Our approach is related to the theory of ergodicity. Ergodic systems are dynamical systems with the property that time averages of functions along trajectories are equal to spatial averages; the associated statistical distributions are known as invariant measures of the system~\cite{petersen_ergodic}. In this work, we construct an ergodic dynamical system where one can prescribe the statistical distributions of the underlying dynamics. By prescribing the color distributions as target distributions for the ergodic system, individual trajectories for each color (the empirical distributions) converge to the desired invariant distributions, thus tracing out the original painting or picture.

The underlying dynamical system is chaotic in the sense that it exhibits sensitivity to initial conditions (as shown in the supplementary material, the system has three Lyapunov exponents $> 0$), but nonetheless leads to robust recreation of the picture irrespective of initial condition. Note that positive Lyapunov exponents are the primary characteristic of chaos~\cite{Cit:Strogatz,Cit:Lyap1}. This algorithm can potentially be used to drive a robot that reproduces paintings/pictures. The ramifications of this approach extend beyond the applicability of designing robotic systems that can paint~\cite{cit:robotPainting,cit:robotPainting1}. The challenging task of efficient sampling of complex probability distributions lies at the heart of a wide range of problems. For example, sampling probability distributions is one of the most important tasks in statistical inference and machine learning, and is typically achieved by Markov Chain Monte Carlo (MCMC) methods~\cite{Cit:Gibbs,Cit:MCMC1,Cit:MCMC2,Cit:MCMC3,Cit:MCMC4}. Variational methods~\cite{Cit:Var1, Cit:Var2} are a popular alternative for statistical inference that rely on the construction of bounds on the likelihood function that may not always be tight~\cite{Cit:Var3}. In this work, we restrict ourselves to MCMC based sampling approaches. MCMC methods are often plagued by slow mixing~\cite{Cit:MCMC_slow}, particularly when distributions are complex and multi-modal. We believe that our approach presents an exciting alternative for sampling complex distributions for Bayesian inference and machine learning in the big data setting. In the supplementary material accompanying this manuscript, we present comparisons of Metropolis-Hastings~\cite{Cit:MCMC_MH}, Hamiltonian MCMC~\cite{Cit:HMC2,Cit:HMC}, and slice sampling~\cite{Cit:Slice} with chaotic sampling. We find that our approach provides higher accuracy and faster computation than all three methods in low dimensions. Additionally, the equations in our approach are easy to construct and can be evolved using Euler or Runge-Kutta integration schemes.

Beyond machine learning and statistical inference, the ability to design dynamical systems with desired properties has a multitude of practical applications. With the emergence of 3-D printing~\cite{Cit:3dprinting} as an approach for quick prototyping of new mechanical parts, the task of achieving desired material distribution becomes increasingly important in aerospace and automobile applications. We imagine that our algorithm can potentially be used to print non-trivial parts, with desired space averaged material properties. This can be achieved by precisely controlling the continuous motion of the printing head using a dynamical system whose trajectory has the desired time averaged statistical properties. Furthermore, we envision that our approach will be used to design coordinated robotic systems that mimic biological swarms for information collection~\cite{cit:fish} (where the statistical time averaged distributions of the trajectories matches the distributions of expected information), construct bio-molecules and associated models with desired statistical behavior or conformations~\cite{cit:DNA,Cit:DNA2}, and build ergodic micro-mixers with optimal mixing properties~\cite{cit:Microfluidics}. The task of analyzing and developing stem cells with appropriate properties of proliferation and differentiation can be modeled using dynamical systems~\cite{Cit:ergodic_stem_cells}. Thus, one can potentially use the construction of appropriate dynamical systems to control gene expressions~\cite{Cit:ergodic_stem_cells} and endow stem cells with the desired statistical proliferation and differentiation properties.

Mathematically, the `painting' problem described above reduces to the problem of designing a continuous time dynamical system whose states sample an arbitrarily complex probability distribution defined over the state space. Leveraging the algorithms described in~\cite{smc_physd}, we construct a dynamical system that `paints' the desired picture. The approach works by decomposing the given picture into its red, green, and blue components, denoted by $\mu^{R}$, $\mu^{G}$, and $\mu^{B}$ respectively. Note that each one of the $\mu$'s captures the spatial distribution of the corresponding color over the entire picture. We then construct three dynamical systems, one for each one of the color distributions, such that the states of these dynamical systems, denoted by $\vec{x}^R$, $\vec{x}^G$, and $\vec{x}^B$, correspond to the two dimensional locations for the red, green, and blue paintbrushes respectively.

To recreate paintings, we would like to design trajectories such that $\vec{x}^{R}$ is ergodic with respect to the distribution $\mu^{R}$, $\vec{x}^{G}$ is ergodic with respect to the distribution $\mu^{G}$ and so on. In other words, we would like to construct ergodic dynamical systems~\cite{cit:katok,cit:equil} with invariant measures given by $\mu^{R}$, $\mu^{G}$, and $\mu^{B}$. Thus, for each color we construct a separate dynamical system, as described below. To sample a given probability distribution $\mu$, we use the notion of a coverage distribution whose support is the set of points in the state space that have already been visited by the generated trajectory. This is defined as:
\begin{equation}
C(\vec{p}) = \frac{1}{ t} \int_{0}^{t} \delta_{\vec{x}(\tau)} (\vec{p}) d \tau,
\label{equ:covdist}
\end{equation}
where $\vec{p}$ is a point in $\mathbb{R}^2$, $t$ is the time, $\delta$ is the Dirac delta function, and $\vec{x}$ is the trajectory for a single color. We would now like the distributions $C$ to ``weakly'' converge to the distributions  $\mu$ as $t\rightarrow\infty$. The difference between the coverage distribution $C$ and $\mu$ is denoted by $\phi(t)$, and is defined as,
\begin{equation}
\phi^2(t) = \| C - \mu \|^2_{H^{-3/2}}\,,
\label{eq:phi}
\end{equation}
where $H^{-3/2}$ denotes the negative Sobolev space norm that captures how close $C$ and $\mu$ are in a ``weak'' sense. This is equivalent to minimizing the difference between the weighted Fourier expansions of both $C$ and $\mu$. Thus, the metric $\phi(t)$ on a two dimensional rectangular domain is computed using,
\begin{eqnarray}
\phi^2(t) &=& \sum_{k} \Lambda_{k} | c_k(t) - \mu_k|^2,\label{eq:phi}\\
\mathrm{where}, &&\nonumber\\
\Lambda_{k} &=& \frac{1}{\left(1 + \|k \|^2 \right)^{3/2}},\label{eq:lam}\quad f_k(x,y) = \frac{1}{h_k}\cos( \frac{k_{x}\pi x}{L_{x}}) \cos(\frac{k_{y}\pi  y}{L_{y}}),\\
c_k(t) &=&  \left < C , f_{k} \right >,  \text{ and } \mu_k = \left < \mu, f_{k}\right>,
\label{eq:main}
\end{eqnarray}
and we take $\vec{x} = [x,y]$, $k=[k_x,k_y]$ is the corresponding wave-number vector, $\left[L_{x},L_{y}\right]$ are the dimensions of the painting or picture, and $<.,.>$ denotes the standard inner product between functions. By minimizing an appropriate function of $\phi(t)$~\cite{smc_physd}, one essentially forces the Fourier coefficients of $C$ (denoted by $c_{k}$) to converge to the Fourier coefficients of $\mu$ (denoted by $\mu_{k}$), but with greater importance given to the large-scale modes (captured by $\Lambda_{k}$). The dynamical system that achieves this minimization is,
\begin{align}
\left[\dot x,\,\, \dot y\right]&= [u_x(t), \,\, u_y(t)], \label{eq:firstorder}\\
\mathrm{where} \quad \left[u_x(t), \,\, u_y(t)\right]&= -u_{\text{max}}  \frac{\left[B_x(t),\,\, B_y(t)\right]}{ ||\left[B_x(t),\,\, B_y(t)\right]||_2} \\
\mathrm{and} \quad \left[B_x(t),\,\, B_y(t)\right]&=  \sum_{k}\Lambda_k t(c_k(t) - \mu_k) \left[ \frac{\partial f_{k}}{\partial x}, \,\, \frac{\partial f_{k}}{\partial y}\right].\label{eq:covcntrl}
\end{align}
Here $u_{\text{max}}$ is the maximum speed of the paintbrushes. $\Lambda_{k}$ serves as a weighting factor that gives greater importance to the large scale features than the finer ones. Of course, the number of Fourier terms has to be truncated to a fixed value $K=K_{x}\times K_{y}$, where we assume that the maximum value for $k_x$  is $K_{x}$ and $k_y$ is $K_{y}$. The higher the value of $K$, the more detailed are the features of the reproduced painting. Additionally, since $C\rightarrow\mu$ in a ``weak'' sense as $t\rightarrow\infty$, the longer one runs the simulation the ``closer'' is the reproduced image to the original. Thus, the reader may be tempted to pick large values for $K$ and $T$ (time for simulation), however, it is important to note that computational burden scales as $O(P\log(P) + KT)$, where $P$ is the number of pixels. Note that the $O(P\log(P))$ complexity arises due to the fast Fourier transform of $\mu$ in Eq.~\ref{eq:main}. For more information about the underlying mathematical theory and extensions to higher dimensions, we refer the reader to the supplementary material. Also, note that the resulting dynamical system described in Eqs~(\ref{eq:firstorder}-\ref{eq:covcntrl}) is chaotic (please refer to the supplementary material).

The main steps of our algorithm can be summarized as follows:
\begin{enumerate}
\item The original image is decomposed into its red, green, and blue components yielding $\mu^{R}$, $\mu^{G}$, and $\mu^{B}$ respectively.
\item Fourier coefficients (based on a preselected value of $K$) of the three color distributions obtained in the previous step are computed.
\item For each one of the colors, the dynamical system described by Eqs~(\ref{eq:firstorder}-\ref{eq:covcntrl}) is evolved for a prescribed amount of time $T$. The fraction of time that each trajectory spends in a pixel determines the intensity of the corresponding color for that pixel.
\end{enumerate}

\begin{figure}[htb!]
    \begin{center}
    \subfigure[Time = $0.016$ sec]{\includegraphics[width=0.51\textwidth]{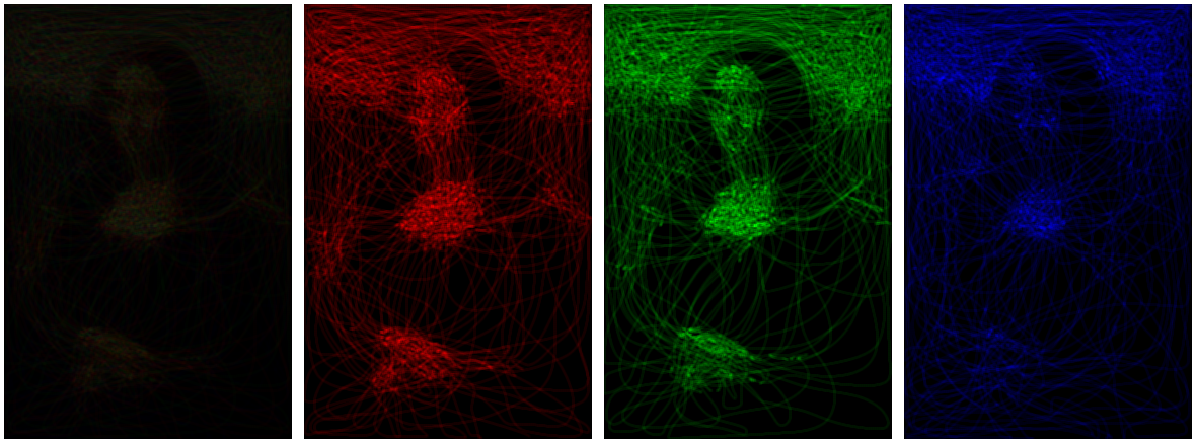}}
    \subfigure[Time = $0.046$ sec]{\includegraphics[width=0.51\textwidth]{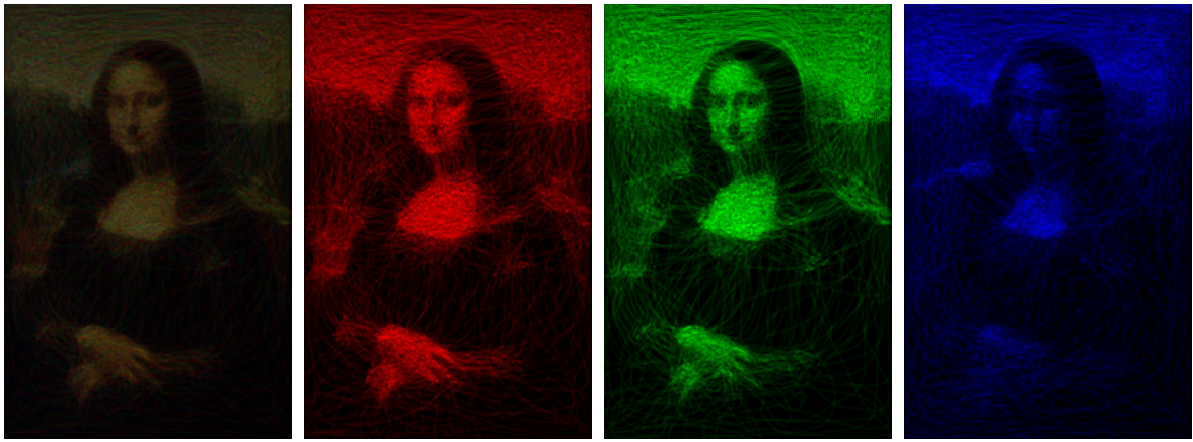}}
    \subfigure[Time = $0.076$ sec]{\includegraphics[width=0.51\textwidth]{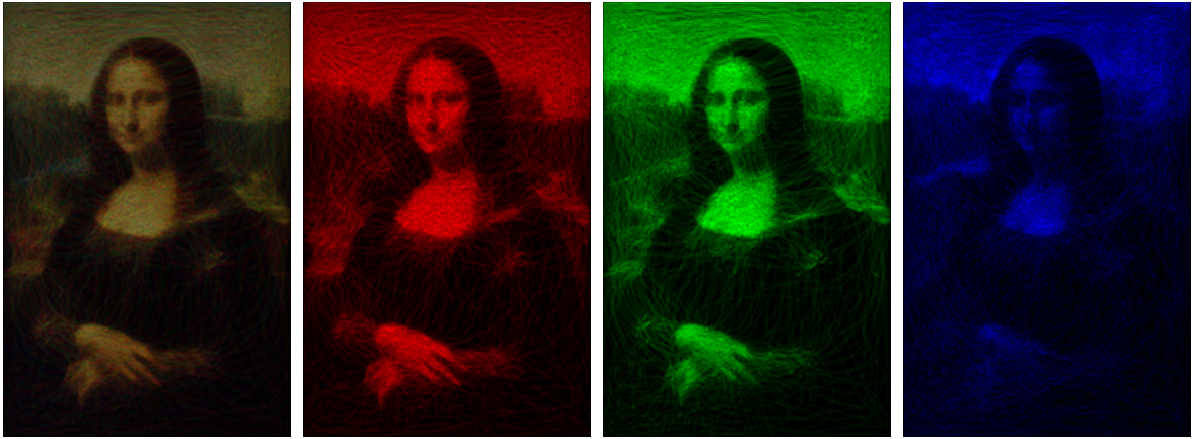}}
    \subfigure[Time = $0.106$ sec]{\includegraphics[width=0.51\textwidth]{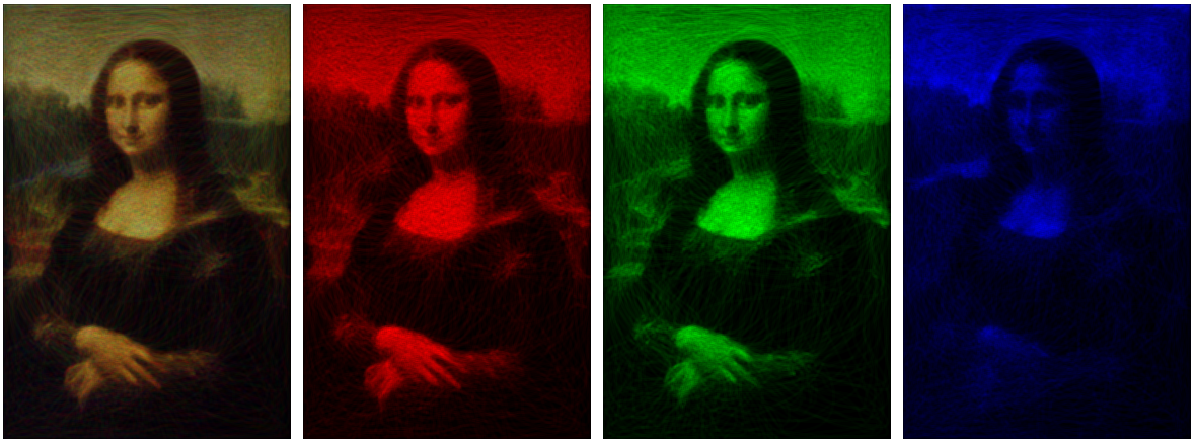}}
    \subfigure[Time = $0.151$ sec]{\includegraphics[width=0.55\textwidth]{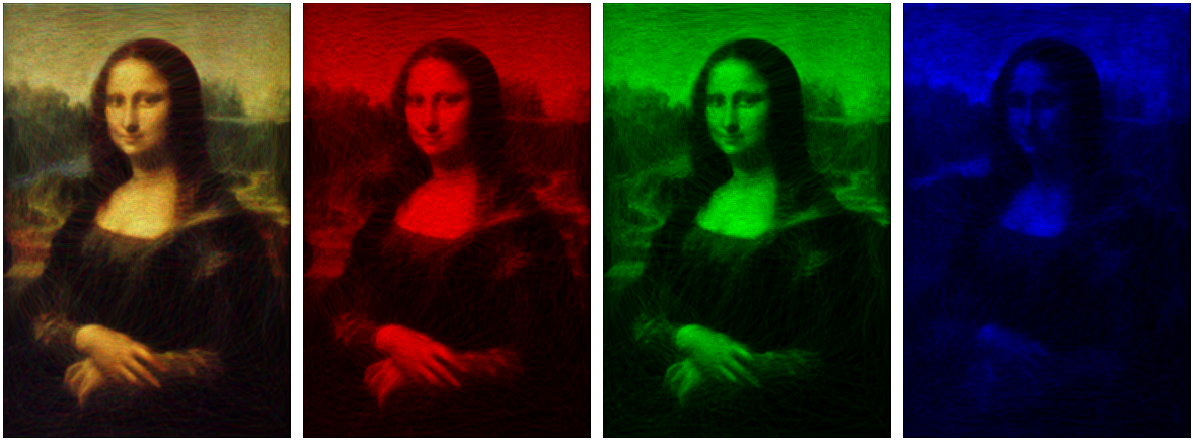}}
    \end{center}
\vspace*{-0.33in}
\caption{Evolving reproduction of the Mona Lisa as recreated by chaotic sampling. The first frame is the superposition of the red, green, and blue frames. Note that the red, green, and blue frames are composed of a single trajectory for each color evolving over time.}
\label{Fig:ergodicmonalisa}
\end{figure}

To demonstrate our algorithm we use Leonardo Da Vinci's iconic painting -  the Mona Lisa. Figure \ref{Fig:ergodicmonalisa} shows the evolution of the Mona Lisa as generated by chaotic sampling. The trajectories of the red, green, and blue paintbrushes are shown in Figure \ref{Fig:ergodicmonalisa}. Here the red, green, and blue images are generated by a single trajectory for each color, this is akin to a continuous motion for each paintbrush that eventually produces the desired distribution of color. The picture is reproduced by just three individual trajectories, one corresponding to each color. As the computation progresses, the Mona Lisa image emerges on the superimposition of the three trajectories. Note that to evolve the equations in~\ref{eq:firstorder}, we use a simple explicit Euler scheme with a step size of $dt=10^{-3}$. The associated movie (Movie S1) displays the remarkable evolution of the reproduction of the Mona Lisa.

\begin{figure}[htb!]
    \begin{center}
    \subfigure[Original Big Sur Photograph]{\includegraphics[width=0.4\textwidth]{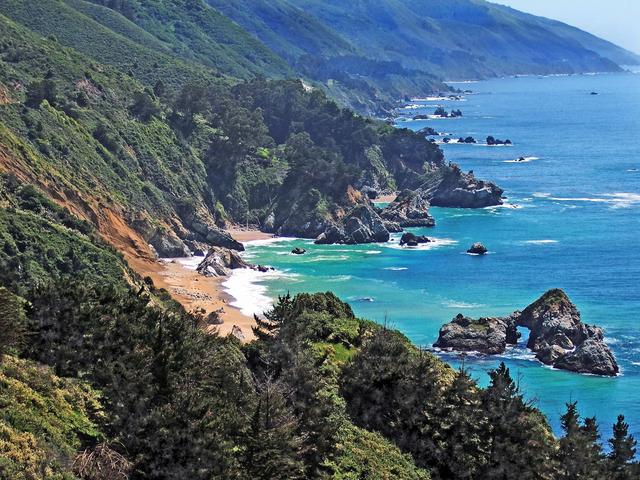}}\hspace{0.5in}
    \subfigure[Big Sur reproduced using chaotic sampling]{\includegraphics[width=0.4\textwidth]{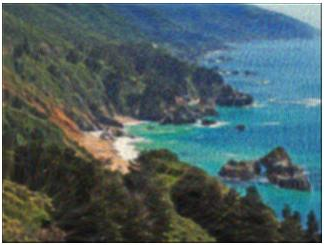}}\\
    \subfigure[Original Pines Switz Photograph]{\includegraphics[width=0.4\textwidth]{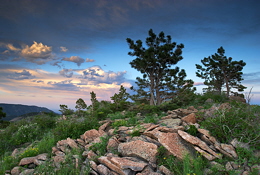}}\hspace{0.5in}
    \subfigure[Pines Switz reproduced using chaotic sampling]{\includegraphics[width=0.4\textwidth]{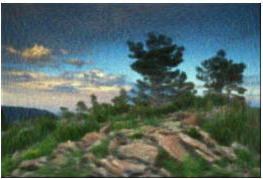}}\\
    \subfigure[Original Starry Night by Van Gogh]{\includegraphics[width=0.4\textwidth]{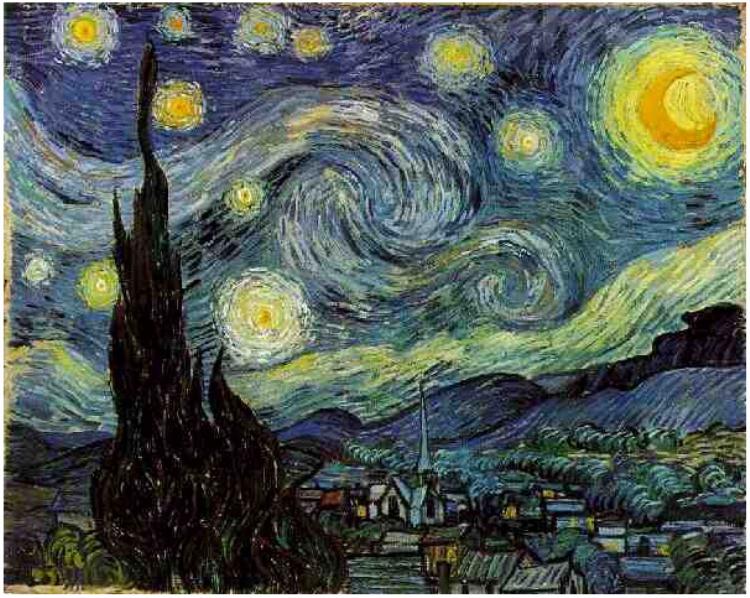}}\hspace{0.5in}
    \subfigure[Starry Night reproduced using chaotic sampling]{\includegraphics[width=0.4\textwidth]{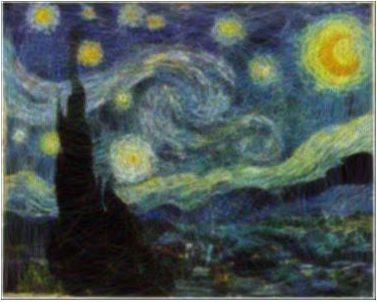}}\\
    \end{center}
\caption{Various pictures and paintings generated by chaotic sampling.}
\label{Fig:ergodicpaintings}
\end{figure}

Figure \ref{Fig:ergodicpaintings} shows various pictures and paintings along with the corresponding reproductions by chaotic sampling. As seen in the cases of Big Sur and Pines Switz, even though the original pictures are real photographs, the reproductions by our dynamical system look surprisingly similar to human paintings. For more information, we refer the reader to Figs. S4-S6 and Movies S2-S4 in the supplementary material.

Note that a host of Markov Chain Monte Carlo (MCMC) techniques for sampling distributions have been developed over the years -  particularly for Bayesian inference and machine learning~\cite{Cit:MCMC3,Cit:MCMC4}. In fact, MCMC methods are a critical step in various statistical and machine learning approaches; thus, these methods form the basis of a very active research community. We compare our chaotic sampling methodology (for $[K_x,K_y] = [80,80]$) with Metropolis-Hastings~\cite{Cit:MCMC_MH, Cit:MH}, Hamiltonian MCMC~\cite{Cit:HMC} and slice sampling~\cite{Cit:Slice} (popular methods for sampling distributions in a wide variety of applications). We use all three methods to sample a multi-modal distribution in two dimensions and find that in comparison to these methods, chaotic sampling provides higher accuracy with faster speeds of computation (see supplementary material for further details). Note that chaotic sampling is not based on constructing Markov chains and in this way is fundamentally different from traditional approaches. Unlike traditional Markov chains that are based on the last sampled point, successive points in chaotic sampling are picked based on the entire history of trajectories (see supplementary material).

Our exposition in the text restricts chaotic sampling to two dimensions, however, there is no such restriction. One can construct a dynamical system to sample probability distributions in any dimension $d$. The general, $d$-dimensional, formulation of chaotic sampling is discussed in the supplementary material. However, the size of the underlying dynamical system in chaotic sampling explodes as $K_{x_1}\times K_{x_2}\times\hdots K_{x_d}$, where $K_{x_i}$ is the wave number in each direction. Our current efforts are focused around addressing this undesirable scaling of the chaotic sampling methodology.

\section*{Conclusions}
In this work, we have developed an algorithmic approach to construct dynamical systems with prescribed statistical properties. We demonstrate that our approach can be used to design chaotic dynamical systems that reproduce paintings and photographs. Akin to a human painter, the dynamical system first captures the large scale features and then fills in the finer details. The given picture is decomposed into its color components, thus yielding distributions of red, green, and blue (or equivalently cyan, magenta, yellow, and key) colors. The algorithm then constructs a separate dynamical system for each color that optimally samples the corresponding color distribution. In a robotic system, these dynamical systems will provide the instructions for each paintbrush with associated colors. These dynamical systems statistically sample the prescribed distributions, consequently, the results are independent of initial conditions. The resulting equations for chaotic sampling are shown (in the supplementary material) to have three positive Lyapunov exponents, implying sensitive dependence to initial conditions, a key property of chaotic systems~\cite{Cit:Strogatz}. The paintings are the ``attractors'' for this dynamical system.

Additionally, in our supplementary material, we investigate the utility of the chaotic sampling approach for machine learning and Bayesian inference in big data settings. In particular, we demonstrate significant gains (in accuracy and convergence time) over traditional MCMC~\cite{Cit:MCMC3} methods. We compare chaotic sampling to slice sampling, Hamiltonian MCMC, and Metropolis-Hastings on a multi-modal test example in two dimensions. Our approach has the advantage over Metropolis-Hastings that it does not require the construction of proposal distributions. The construction of these distributions can be challenging. Moreover, given that the problem of designing systems with prescribed statistical properties arises in numerous applications such as 3-D printing~\cite{Cit:3dprinting}, biological systems~\cite{cit:DNA,Cit:DNA2} and microfluidics~\cite{cit:Microfluidics}, we anticipate our approach will also be valuable in these scenarios.

\bibliography{ErgodicPainting}
\bibliographystyle{Science}


%

\clearpage

\section*{Supplementary Text}

\subsection*{Ergodic Trajectory Generation}
In this section, we construct dynamical systems that sample prescribed distributions. Let the dynamical system  trajectory be denoted by $\vec{x}(t) \in U \subset \mathbb{R}^{d}$. Note that in the case of paintings $d=2$ (as paintings and pictures are two dimensional). To keep track of the points already visited by the dynamical system, we define the coverage distribution (at location $\vec{p}\in \mathbb{R}^d$), generated by the trajectory $\vec{x}$ as,
\begin{equation}
C(\vec{p}) = \frac{1}{ t} \int_{0}^{t} \delta_{\vec{x}(\tau)} (\vec{p}) d \tau,
\label{equ:covdist}
\end{equation}
where $t$ is the time and $\delta$ is the Dirac delta measure. For the coverage distribution defined above, let us compute its spherical integrals given as:
\begin{equation}
d(\vec{p}, r) = \langle C, \chi_{B(\vec{p},r)} \rangle = \int\limits_{B(\vec{p},r)} C(\vec{z}) d\vec{z}.
\end{equation}
where $B(\vec{p},r) = \{\vec{z}: \|\vec{z} - \vec{p} \| \leq r \}$ and $\chi_{B(\vec{p},r)}$ is the indicator function on the set $B(\vec{p},r)$. The resulting spherical integral $d(\vec{p},r)$ has a very useful interpretation; $d(\vec{p},r)$ is the fraction of time spent by the trajectory in the set $B(\vec{p},r)$.

The coverage metric we use is a distance between the coverage distribution and a probability distribution $\mu$. Here $\mu$ represents the significance (or importance) that should be given to each individual point in the domain. In other words, let $\mu$ be the target distribution defined over a region $U \subset\mathbb{R}^d$. Consider a distance given by comparing the differences in the spherical integrals of $C$ and $\mu$:
\begin{equation}\label{cost1}
E^2(t) = \int_{0}^{R} \int \limits_{U} \left( d(\vec{p}, r) - \mu(B(\vec{p},r)) \right)^2 d\vec{p}\, dr.
\end{equation}
Also consider the distance given by the Sobolev space norm which can be expressed as,
\begin{equation}
\begin{split}
\phi^2(t) = \| C(.) - \mu(.) \|^2_{H^{-3/2}} &= \sum_{K \in {\mathbb{Z}^*}^d} \Lambda_{k} | c_k(t) - \mu_k|^2,  \\
\text{where } \Lambda_{k} = \frac{1}{\left(1 + \|k \|^2 \right)^{(d+1)/2}}, \,\,
c_k(t) =  \left < C , f_k \right > & \text{ and } \mu_k = \left < \mu, f_k \right>.
\end{split}
\end{equation}
 Here $\{f_k\}$ are the $d$-dimensional Fourier basis functions that satisfy Neumann boundary conditions on the domain $U$, $k=[k_{x_1},k_{x_2},\dots,k_{x_d}]$ is the corresponding wave-number vector which belongs to ${\mathbb{Z}^*}^d =\left[0,1,2,\hdots\right]^{d}$. For instance, on a rectangular domain $U = [0, L_{x}] \times [0, L_{y}]$, $\vec{x} = [x,y]$, and $k=[k_{x},k_{y}]$. This gives,
\begin{equation}
\begin{split}
f_k(x,y) &= \frac{1}{h_k}\cos( \frac{k_{x}\pi x}{L_{x}}) \cos(\frac{k_{y}\pi  y}{L_{y}}), \text{where } k_{x}, k_{y} = 0,1,2.... \\
\end{split}
\label{equ:neumannbasis}
\end{equation}
where $h_k$ is the normalization constant. The two metrics $E$ and $\phi$ described above are equivalent, i.e., there exist bounded constants $c_1, c_2$ such that,
\begin{equation}
c_{1} \phi^2 \leq E^2 \leq c_{2} \phi^2.
\end{equation}
For more details on the relationship of these metrics to concepts in ergodic theory, we refer the reader to $\citesmcphysd$ . Since the metric $\phi$ is easier to compute, we use $\phi$ for control design. We aim to design a control such that $\phi^{2}(t)\rightarrow 0$. This ensures that the distribution $C(.)\rightarrow \mu(.)$ as $t\rightarrow\infty$.

Assume that the control $\vec{u}(t)$ describes the sampling trajectory for the system,
\begin{equation}
\dot{\vec{x}}(t) = \vec{u}(t).
\label{equ:firstorder}
\end{equation}
Let us define the following vector for each dynamical system:
\begin{equation}
\begin{split}
\vec{B}(t) &= \sum_{k} \Lambda_k s_k(t) \vec{\nabla} f_k (\vec{x}(t)).
\end{split}
\end{equation}
Here, $s_{k} = t(c_{k}(t) - \mu_{k})$ and $\vec{\nabla} f_k (\vec{x}(t))$ is the gradient of the Fourier basis function evaluated at state $\vec{x}(t)$. The coverage control is given as:
\begin{equation}
\vec{u}(t) =
-u_{\text{max}}  \frac{\vec{B}(t)}{ \| \vec{B}(t) \|_2},
\label{equ:covcntrl}
\end{equation}
where $u_{\text{max}}$ is the maximum rate of evolution of each dynamical system.

Thus, the overall equations (for each color or dynamical system) are given by,
\begin{align}
\dot{\vec{x}}(t) &= \vec{u}(t),\nonumber\\
\vec{u}(t) &= -u_{\text{max}}  \frac{\vec{B}(t)}{ \| \vec{B}(t) \|_2},\nonumber\\
\vec{B}(t) &= \sum_{k} \Lambda_k s_k(t) \vec{\nabla} f_k (\vec{x}(t)),\nonumber\\
s_{k} &= t(c_{k}(t) - \mu_{k}), \nonumber\\
\mu_{k} &= \langle \mu,f_{k} \rangle,\nonumber\\
c_{k} &= \langle C,f_{k} \rangle = \frac{1}{t}\int_{0}^{t} f_{k}(\vec{x}(\tau))d\tau.
\label{eq:final}
\end{align}

The paintings and pictures are reproduced by evolving the above equations in two dimensions and setting the color distributions to $\mu$. The trajectories for each color are then superimposed to produce the overall painting or picture. Note that the computation time for sampling a single $\mu$ can potentially be accelerated by computing, in parallel, multiple trajectories starting from different initial conditions and superimposing the results.
\subsection*{Lyapunov Exponents of the Dynamical System}
One of the key signatures of chaos is the sensitive dependence of the underlying dynamics to initial conditions $\citeStrogatz$. The Lyapunov spectrum $\citeLyapStrogatz$ is defined as,
\begin{align}
\lambda_{i} = \lim_{T\rightarrow \infty}\frac{1}{T}\log(\frac{\Delta_{i}(T)}{\Delta_{i}(0)}),
\label{eq:lyap}
\end{align}
where $\Delta_{i}(0)$ is the initial perturbation of the trajectory along the $i$-th principal axis. Similarly, $\Delta_{i}(T)$ is the divergence of the trajectory from the original trajectory, at time $T$, along the $i$-th principal axis. Consider a two dimensional ($\vec{x}=[x,y]$) rectangular region of dimensions $\left[L_{x},L_{y}\right]$ with a uniform prior (equivalent to a picture with identical pixels). To compute the Lyapunov exponents we use the approach outlined in $\citeLyap$. The resulting equations are given below,
\begin{align}
\dot{x}(t) &= -u_{\text{max}}\frac{B_{x}}{B_{x}^{2} + B_{y}^{2}},\nonumber\\
\dot{y}(t) &= -u_{\text{max}}\frac{B_{y}}{B_{x}^{2} + B_{y}^{2}},\nonumber\\
\dot s_{k}(t) &= f_{k} - \mu_{k},\nonumber \\
\textrm{where},\nonumber\\
B_{x}(t) &= \sum_{k} \Lambda_k s_k(t) \frac{\partial f_{k} (x,y)}{\partial x},\nonumber\\
B_{y}(t) &= \sum_{k} \Lambda_k s_k(t) \frac{\partial f_{k} (x,y)}{\partial y},
\label{eq:lyap_full}
\end{align}
where $f_{k}$ are defined in Eqn.~\ref{equ:neumannbasis}. Note that for accurate computation of Lyapunov exponents, the equations for the entries of the Jacobian are typically included $\citeLyap$. For the above equations, one can derive analytical expressions for the entries of the Jacobian, given by,
\begin{align}
J = \begin{bmatrix}\frac{\partial\dot x}{\partial x} & \frac{\partial\dot x}{\partial y} & \dots & \frac{\partial\dot x}{\partial s_{k}} & \dots \\
\\
\frac{\partial\dot y}{\partial x} & \frac{\partial\dot y}{\partial y} & \dots & \frac{\partial\dot y}{\partial s_{k}} & \dots \\
\vdots & \vdots & & \vdots & \vdots \\
\frac{\partial\dot s_{k}}{\partial x} & \frac{\partial\dot s_{k}}{\partial y} & \dots & \frac{\partial\dot s_{k}}{\partial s_{k}} & \dots
\end{bmatrix},
\end{align}
where,
\begin{align}
\frac{\partial\dot x}{\partial (.)}     &= -u_{\text{max}}B_{y}\frac{\left[B_{y}B_{x}^{'} - B_{x}B_{y}^{'}\right]}{(B_{x}^{2} + B_{y}^{2})^{\frac{3}{2}}},\nonumber\\
\frac{\partial\dot y}{\partial (.)}     &= -u_{\text{max}}B_{x}\frac{\left[B_{x}B_{y}^{'} - B_{y}B_{x}^{'}\right]}{(B_{x}^{2} + B_{y}^{2})^{\frac{3}{2}}},\nonumber\\
\frac{\partial\dot s_{k}}{\partial x} &= \frac{\partial f_{k}}{\partial x},\nonumber\\
\frac{\partial\dot s_{k}}{\partial y} &= \frac{\partial f_{k}}{\partial y},\nonumber\\
\frac{\partial\dot s_{k}}{\partial s_{k}} &= 0.
\end{align}
Thus, by analytically calculating $B_{x}^{'}$ and $B_{y}^{'}$ with respect to $x,y$ and $s_{k}$, one can compute the dynamics of the Jacobian. The dimensionality of this system of equations depends on the number of wave functions that are included in the expansion. In particular, the dynamics of $x,y$ and $s_{k}$ give rise to $M=K_{x}K_{y} + 2$ equations, where $K_x$ and $K_{y}$ are the maximum values for $k_{x}$ and $k_{y}$ respectively. Consequently, the dynamics of the Jacobian is determined by $M^{2}$ equations.

Using the above equations along with the approach outlined in $\citeLyap$, we can compute the dynamics of the entire spectrum of Lyapunov exponents (as shown in Fig. $\FigLyap$). It can be seen that the system has three positive Lyapunov exponents, where the largest exponent has an asympototic value of $\approx 0.1$. Also, note that the information dimension of the attractor $\citeLyap$ is not defined since $\displaystyle\sum_{i=1}^{N}\lambda_{i}>0$.

Thus, the dynamical system is chaotic since it displays sensitive dependence to initial conditions. Note, however, that the final statistical distribution is invariant and independent of initial conditions.

\subsection*{Comparison of Chaotic sampling with Metropolis-Hastings, Hamiltonian MCMC and Slice Sampling}

In this section, we investigate the use of chaotic sampling to sample distributions for machine learning and statistical applications $\citeMCMCbasic$. Markov Chain Monte Carlo (MCMC) methods are used extensively in the areas of machine learning and Bayesian statistics~\citeStatistics, computational physics, and rare event sampling. Additionally, these methods are extensively used in the big data setting. In fact, MCMC methods are often a critical step in various statistical and machine learning approaches; thus, these methods form the basis of a very active research community.  We compare our chaotic sampling methodology with Metropolis-Hastings $\citeMCMCMH$, Hamiltonian MCMC $\citeHMC$ and slice sampling $\citeSlice$ (popular methods for sampling distributions in a wide variety of applications). We use all three methods to sample a multi-modal distribution in two dimensions and find that compared to competing methods, chaotic sampling provides higher accuracy with faster speeds of computation. As mentioned previously, chaotic sampling is not based on constructing Markov chains and in this way is fundamentally different from traditional methods. Unlike traditional Markov chains that are based on the last sampled point, successive points in chaotic sampling are picked based on the entire history of the trajectories. We now give a brief description of Metropolis-Hastings, Hamiltonian MCMC, and slice sampling, and present comparisons.

Metropolis-Hastings is a popular Markov Chain Monte Carlo approach that proceeds by generating samples using a proposal distribution that are then accepted or rejected based on the target distribution $\citeMH$. Thus, Metropolis-Hastings requires the tuning of a proposal distribution, $Q(\vec{x}_{t+1}|\vec{x}_{t})$, that proposes the new state $\vec{x}_{t+1}$ based on $\vec{x}_{t}$. Once the new point $\vec{x}_{t+1}$ is generated using the proposal distribution, it is accepted or rejected using threshold criteria on the target distribution $\mu(\vec{x})$. The construction of the proposal distribution can be particularly challenging $\citeMCMCMH$, and requires much trial and error. For our approach, we pick,
\begin{align}
Q(\vec{x}_{t+1}|\vec{x}_{t}) = \mathcal{N}(\vec{x}_{t},\sigma),
\label{eq:proposal}
\end{align}
where $\mathcal{N}$ is the Gaussian distribution with the arguments of the mean and standard deviation. We find that a standard deviation of $\sigma=1$ (in both $x$ and $y$ directions) works best for the multi-modal example described later in this section.

The Hamiltonian MCMC $\citeHMCall$ is an approach for sampling distributions that is based on the imposition of a Hamiltonian structure on the underlying dynamics. Hamiltonian MCMC proceeds by considering the original variables or states as the ``position'' variables and appending ``momentum'' variables. The distribution of momentum variables is typically assumed to be Gaussian $\citeHMCO$. One then simulates a Markov chain by resampling the momentum variables and then performing Metropolis updates on the position variables. Note that proposal states in Hamiltonian MCMC are not generated by an explicit proposal distribution, but by the momentum variable updates. The evolution of the dynamics of the Hamiltonian system is typically performed using symplectic integrators $\citeHMCO$. The advantage of Hamiltonian MCMC over Metropolis-Hastings is that correlation between successive samples is avoided by using a Hamiltonian structure on the underlying states $\citeHMCO$. The disadvantages are that the target distribution of the momentum variables can be difficult to design and the number of states have to be doubled due to the Hamiltonian structure (since one must introduce momentum variables).

Slice sampling $\citeSlice$ is based on uniformly sampling the graph of a density function. This is achieved by alternatively sampling (using uniform distributions) the state and probability spaces. Essentially, one uniformly samples points from the vertical interval defined by the density of the current point, followed by uniform sampling of the union of intervals that constitute the horizontal ``slice'' of the density function. The advantages of slice sampling are that one does not require any parameter or proposal density selection and ease of implementation. Our results on slice sampling are based on the default implementation in the MATLAB software package.

To compare our chaotic sampling approach with Metropolis-Hastings $\citeMCMCMH$, Hamiltonian MCMC $\citeHMC$, and slice sampling $\citeSlice$, we pick a multi-modal distribution by normalizing the sum of three Gaussian distributions in two dimensions. The first Gaussian is centered at $\left(-2.0,-2.0\right)$, the second Gaussian at $\left(2.0,2.0\right)$, and the third Gaussian at $\left(-2.0,2.0\right)$. All the Gaussian distributions have a standard deviation of $0.5$ and a correlation $\rho=0$ (see Figure $\Figthreepeaks$). Thus, the probability distribution is given by,
\begin{align}
\mu(x,y) = \frac{1}{6\pi\sigma_{x}\sigma_{y}}\displaystyle\sum_{i=1}^{3}\exp(-\left[\frac{(x-m_{x_{i}})^2}{\sigma_{x}^2} + \frac{(y-m_{y_{i}})^2}{\sigma_{y}^2}\right])
\label{eq:target}
\end{align}

The advantage of picking a distribution of this form is that the mean and higher moments can be computed analytically. For example, the mean of $\mu(x,y)$ is given by $m_{x} = \frac{m_{x_1}+m_{x_2} + m_{x_3}}{3}$ and $m_{y} = \frac{m_{y_1}+m_{y_2} + m_{y_3}}{3}$.  We generate $8000$ samples using the following methods: chaotic sampling, Metropolis-Hastings (with the proposal density given in Eqn.~\ref{eq:proposal}), Hamiltonian MCMC, and slice sampling, treating the distribution in Eqn.~\ref{eq:target} (Figure $\Figthreepeaks$) as the target distribution. We then compute the statistical error of the various approaches (with respect to the analytical closed form solutions) as function of the number of samples.

We can compare the moments of any observable on the $\vec{x}=\left(x,y\right)$ space. For simplicity, we choose $\mathbb{E}(x)$, note however, that any complicated integral can be used for comparison. The results are averaged over $10$ trials (with $8000$ samples each) and the convergence of the error in the predicted mean of $x$ is shown in Figure $\FigMCMC$.

Note that for chaotic sampling we pick $\left[K_x,K_y\right]=\left[80,80\right]$ and $dt=0.1$ for the explicit Euler integration scheme. We generate the samples using the dynamical system described by Eqns.~\ref{equ:firstorder} and~\ref{equ:covcntrl}. We first deterministically run the dynamical system and then, to penalize points on the trajectory that lie between the peaks (since they only serve to connect regions of high probability), we reject the points on the trajectory where $\mu(x,y)$ lies below a threshold (we pick $0.05$ in this case). Note that this point rejection step is performed after the generation of all the points, and has complexity $O(\frac{T}{dt})$.

The comparison of all the methods is presented in Figure $\FigMCMC$. The figure shows the error in the estimate of the mean of the two dimensional multi-modal distribution as a function of the number of samples (the results are averaged over $10$ independent runs). It is clear in Figure $\FigMCMC$ that the chaotic sampling method converges significantly faster than Metropolis-Hastings, Hamiltonian MCMC, and slice sampling. Additionally, we find that for the three peak example, chaotic sampling approach is computationally $1.8$X faster than Metropolis-Hastings, $3$X faster than slice sampling, and $10$X faster than Hamiltonian MCMC. Furthermore, note that the chaotic sampling approach does not require the construction of a proposal distribution that can be complicated $\citeMCMCMH$.

Further analysis of chaotic sampling is required in the context of Markov Chain Monte Carlo (MCMC) sampling. Primarily, we aim to address the requirement of computing Fourier integrals for chaotic sampling as well as constructing ``sparse'' representations of the probability distributions to reduce the number of coefficients in high dimensions. Additionally, we are also investigating the development of ``hybrid'' methods that first use traditional sampling approaches to obtain rough estimates of the Fourier integrals and then switch over to chaotic sampling.

\subsection*{Test Runs of Chaotic Sampling on various Pictures and Paintings}
In addition to the Mona Lisa simulations, we present the evolution of reproductions of various pictures and paintings using chaotic sampling. All simulations were run using explicit Euler integration with $dt=10^{-3}$ and $[K_{x},K_{y}] = [100,100]$. Here we present time snapshots of the reproductions of Pines Switz (Figure $\Figpineswitz$) and Big Sur (Figure $\Figbigsur$) photographs, and the Starry Night painting (Figure $\Figstarrynight$). These results, just as in the Mona Lisa example, are produced by computing a single trajectory for each color. These trajectories are superimposed to produce the overall picture. Additionally, the corresponding time evolution of the reproduction of all the paintings and pictures are captured in Movies S1-S4. One can clearly see the emergence of the pictures and paintings as the computation progresses.



\begin{figure}[htb!]
    \begin{center}
    \includegraphics[scale=0.7]{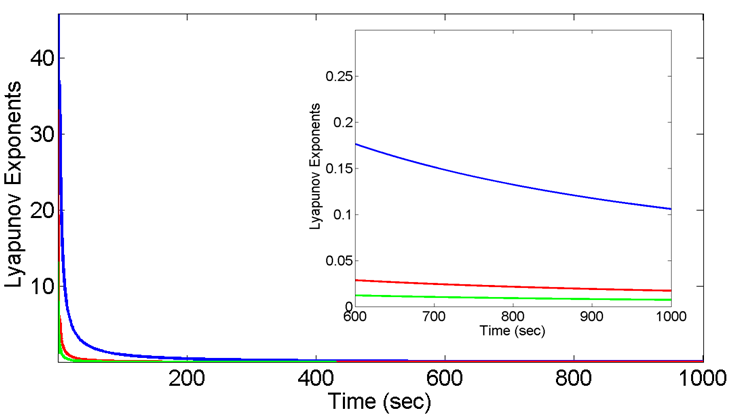}
    \end{center}
\caption*{\textbf{Fig. S1:} A plot of the positive Lyapunov exponents vs. simulation time. The inset figure zooms into the interval $t=[600,1000]$. It can be clearly seen that three Lyapunov exponents are positive, thus implying chaos.}
\label{Fig:Lyap}
\end{figure}

\begin{figure}[htb!]
    \begin{center}
    \includegraphics[scale=0.8]{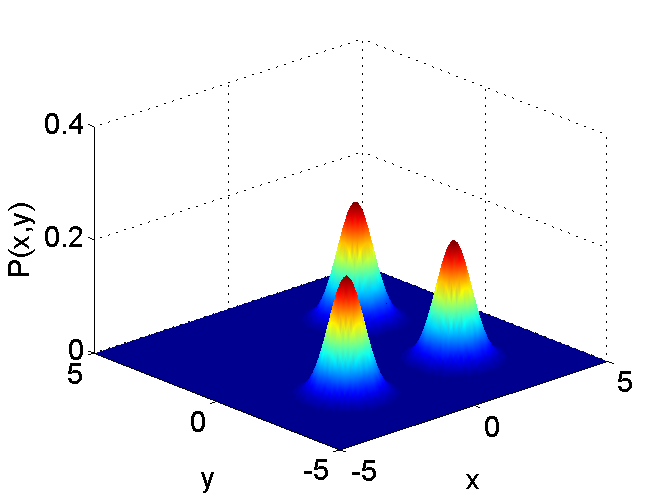}
    \end{center}
\caption*{\textbf{Fig. S2:} The multi-modal distribution used in the comparison of chaotic sampling, Metropolis-Hastings, Hamiltonian MCMC, and slice sampling.}
\label{Fig:three_peaks}
\end{figure}

\begin{figure}[htb!]
    \begin{center}
    \includegraphics[scale=0.4]{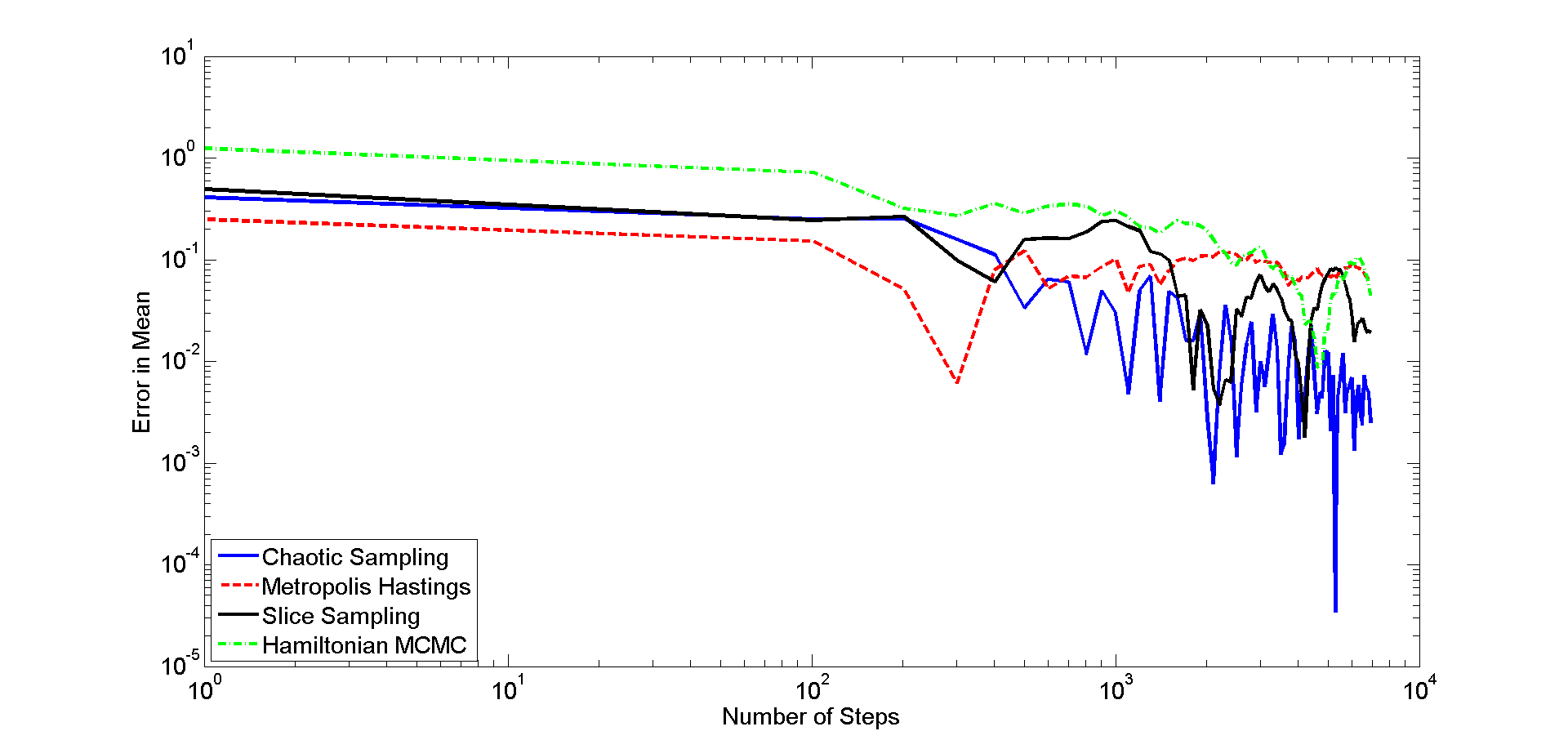}
    \end{center}
\caption*{\textbf{Fig. S3:} Comparison of convergence of chaotic sampling, Metropolis-Hastings, Hamiltonian MCMC, and slice sampling.}
\label{Fig:MCMC}
\end{figure}

\begin{figure}[htb!]
    \begin{center}
    \subfigure[Time = $0.016$ sec]{\includegraphics[width=0.85\textwidth]{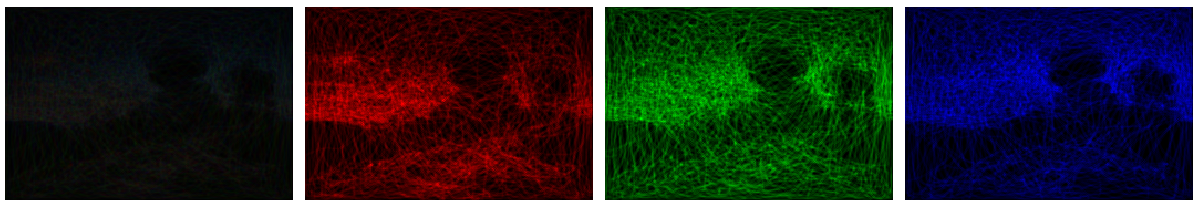}}
    \subfigure[Time = $0.031$ sec]{\includegraphics[width=0.85\textwidth]{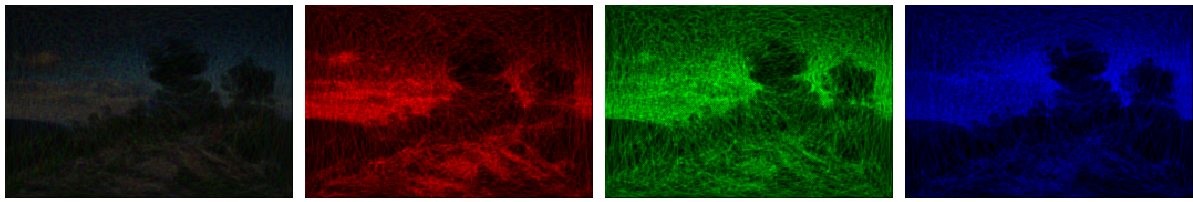}}
    \subfigure[Time = $0.061$ sec]{\includegraphics[width=0.85\textwidth]{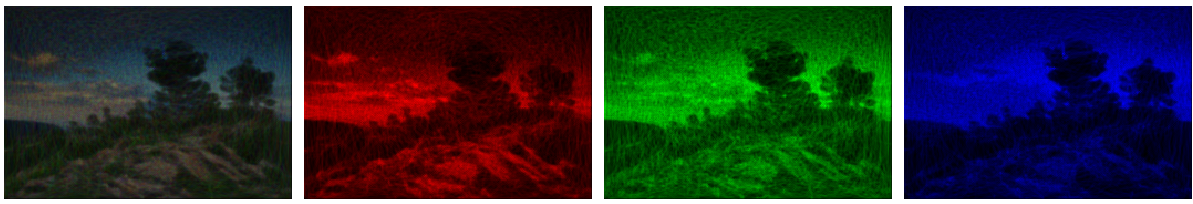}}
    \subfigure[Time = $0.091$ sec]{\includegraphics[width=0.85\textwidth]{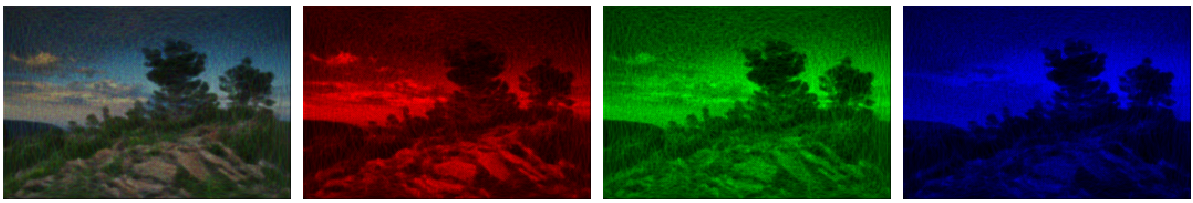}}
    \subfigure[Time = $0.151$ sec]{\includegraphics[width=0.85\textwidth]{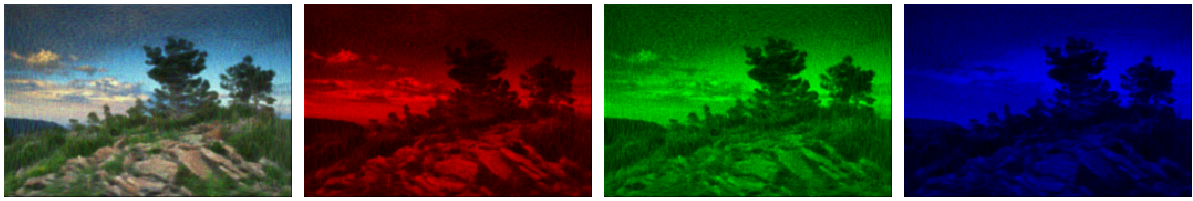}}
    \end{center}
\caption*{\textbf{Fig. S4:} Evolving reproduction of the Pines Switz photograph as recreated by chaotic sampling. The first frame is the superposition of the red, green, and blue frames. Note that the red, green, and blue frames are composed of a single trajectory for each color evolving over time.}
\label{Fig:pineswitz}
\end{figure}

\setcounter{subfigure}{0}
\begin{figure}[htb!]
    \begin{center}
    \subfigure[Time = $0.016$ sec]{\includegraphics[width=0.85\textwidth]{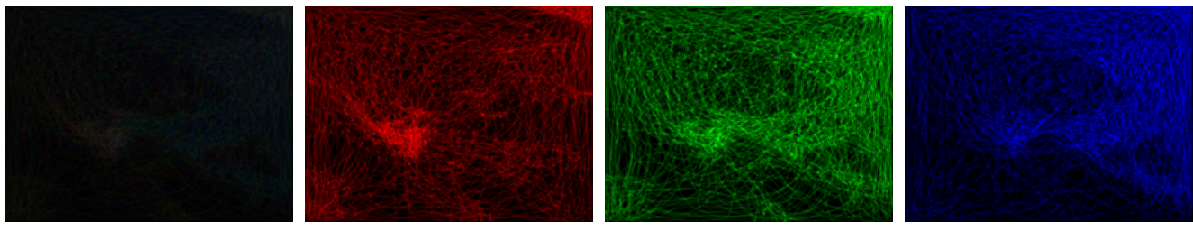}}
    \subfigure[Time = $0.031$ sec]{\includegraphics[width=0.85\textwidth]{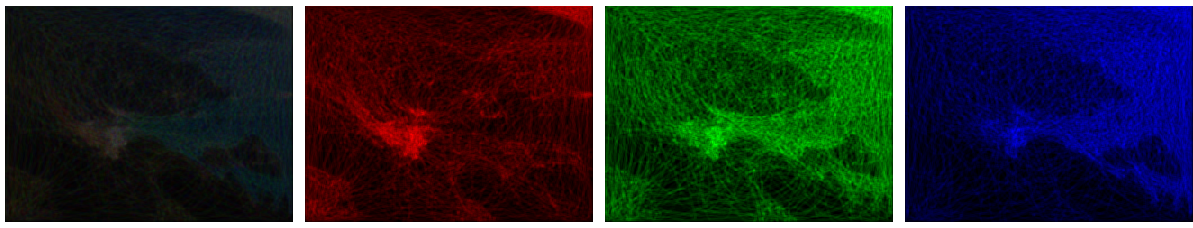}}
    \subfigure[Time = $0.061$ sec]{\includegraphics[width=0.85\textwidth]{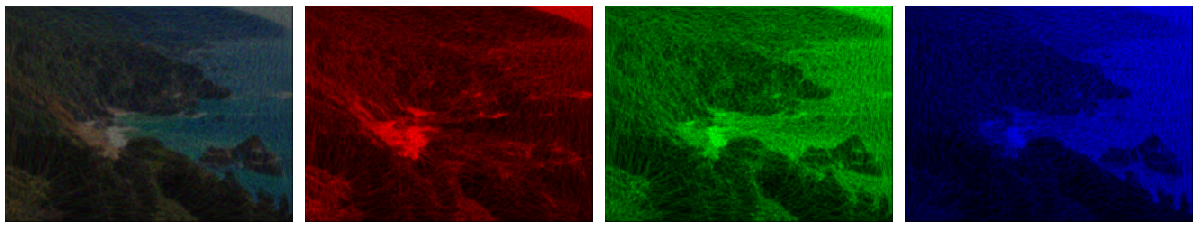}}
    \subfigure[Time = $0.091$ sec]{\includegraphics[width=0.85\textwidth]{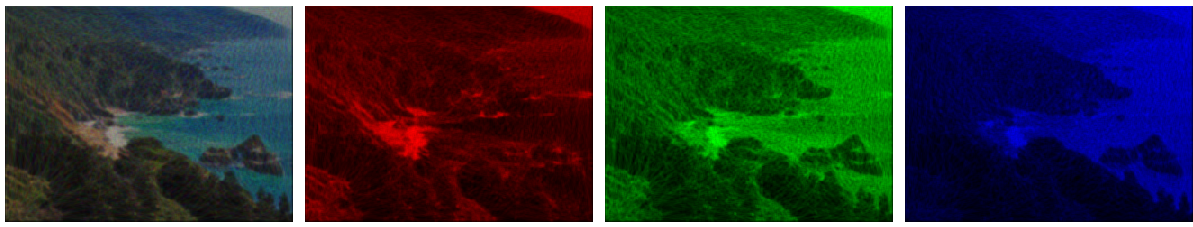}}
    \subfigure[Time = $0.151$ sec]{\includegraphics[width=0.85\textwidth]{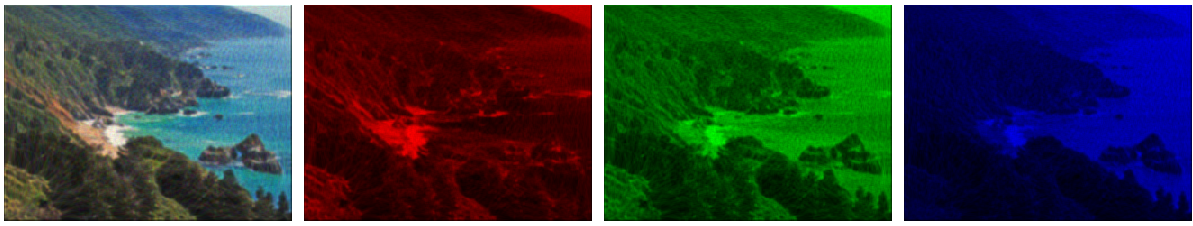}}
    \end{center}
\caption*{\textbf{Fig. S5:} Evolving reproduction of the Big Sur photograph as recreated by chaotic sampling. The first frame is the superposition of the red, green, and blue frames. Note that the red, green, and blue frames are composed of a single trajectory for each color evolving over time.}
\label{Fig:bigsur}
\end{figure}

\setcounter{subfigure}{0}
\begin{figure}[htb!]
    \begin{center}
    \subfigure[Time = $0.016$ sec]{\includegraphics[width=0.85\textwidth]{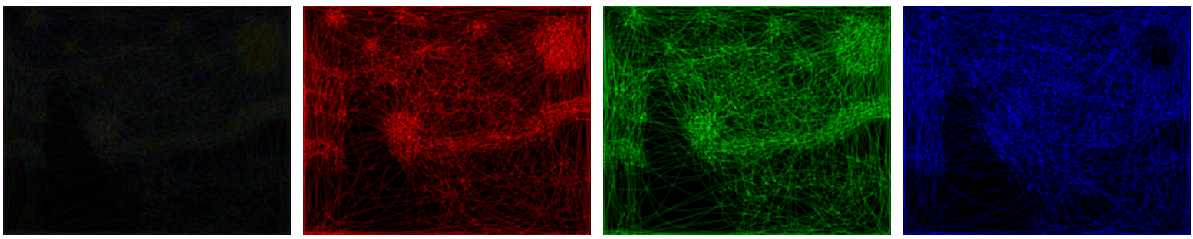}}
    \subfigure[Time = $0.031$ sec]{\includegraphics[width=0.85\textwidth]{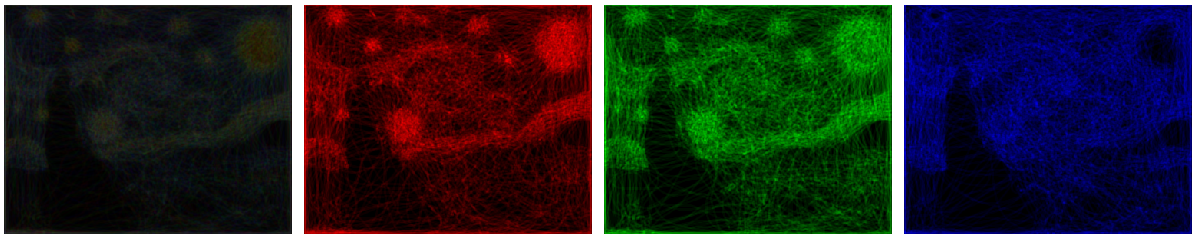}}\\
    \subfigure[Time = $0.061$ sec]{\includegraphics[width=0.85\textwidth]{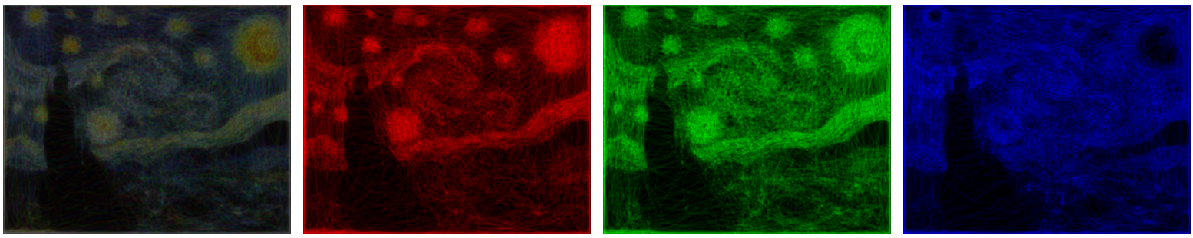}}
    \subfigure[Time = $0.091$ sec]{\includegraphics[width=0.85\textwidth]{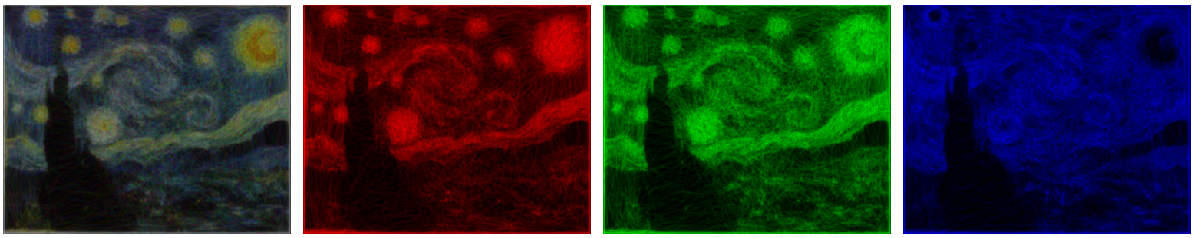}}\\
    \subfigure[Time = $0.151$ sec]{\includegraphics[width=0.85\textwidth]{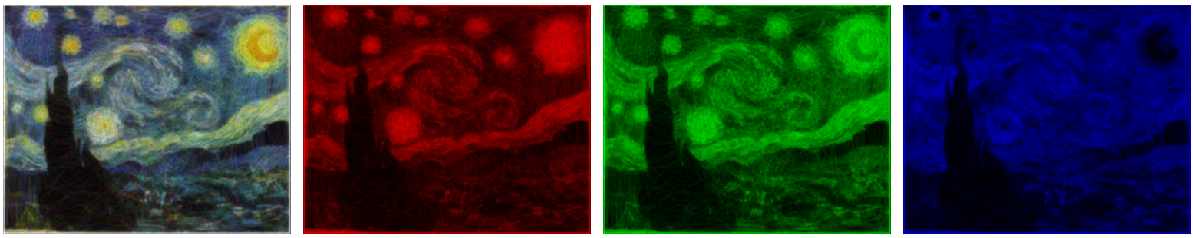}}\\
    \end{center}
\caption*{\textbf{Fig. S6:} Evolving reproduction of the Starry Night painting as recreated by chaotic sampling. The first frame is the superposition of the red, green, and blue frames. Note that the red, green, and blue frames are composed of a single trajectory for each color evolving over time.}
\label{Fig:starrynight}
\end{figure}

\clearpage
\textbf{Movie S1}\\
The movie shows the chaotic sampling based reproduction of the Mona Lisa painting. The first frame is the superposition of the red, green, and blue frames. Note that the red, green, and blue frames are composed of a single trajectory for each color evolving over time.

\textbf{Movie S2}\\
The movie shows the chaotic sampling based reproduction of the Pines Switz photograph. The first frame is the superposition of the red, green, and blue frames. Note that the red, green, and blue frames are composed of a single trajectory for each color evolving over time.

\textbf{Movie S3}\\
The movie shows the chaotic sampling based reproduction of the Big Sur photograph. The first frame is the superposition of the red, green, and blue frames. Note that the red, green, and blue frames are composed of a single trajectory for each color evolving over time.

\textbf{Movie S4}\\
The movie shows the chaotic sampling based reproduction of the Starry Night painting. The first frame is the superposition of the red, green, and blue frames. Note that the red, green, and blue frames are composed of a single trajectory for each color evolving over time.

\end{document}